\documentclass[prl,twocolumn,groupedaddress,showpacs,nofootinbib]{revtex4}
\usepackage{graphicx}
\usepackage{dcolumn}
\usepackage{amssymb}
\usepackage{mathrsfs}
\usepackage{amsmath}
\usepackage{epsfig}
\usepackage[dvips]{color}
\usepackage{hhline}

\begin{document}

\title{Natural Inflation and Flavor Mixing from Peccei-Quinn Symmetry Breaking}

\author{Pei-Hong Gu}
\email{peihong.gu@mpi-hd.mpg.de}

\author{Manfred Lindner}
\email{manfred.lindner@mpi-hd.mpg.de}

\affiliation{Max-Planck-Institut f\"{u}r Kernphysik, Saupfercheckweg
1, 69117 Heidelberg, Germany}

\begin{abstract}

We propose a left-right symmetric model to simultaneously give
natural inflation and flavor mixing from a Peccei-Quinn symmetry
breaking at the Planck scale. Our model can be embedded in
$SO(10)$ grand unification theories.

\end{abstract}


\maketitle

\emph{Introduction}: The Peccei-Quinn \cite{pq1977} (PQ) symmetry
predicts the existence of an axion  \cite{pq1977,weinberg1978} which
would solve the strong CP problem. In addition the axion could be
dark matter and the PQ symmetry could have various other interesting
implications for particle physics and cosmology \cite{kc2008}. For
example \cite{pi1984}, in an $SU(5)$ grand unification theory (GUT),
the PQ symmetry breaking at the Planck scale can result in a pseudo
Nambu-Goldstone boson (pNGB) with a Coleman-Weinberg potential
\cite{cw1973} to drive inflation \cite{guth1981}. One can also
relate the PQ symmetry breaking to the neutrino mass-generation
\cite{shin1987}. In particular, the PQ symmetry can be embedded into
the $SU(3)_c^{}\times SU(2)_L^{}\times S(2)_R^{}\times
U(1)_{B-L}^{}$ left-right symmetric theories \cite{ps1974} if the
left-right symmetry is charge-conjugation \cite{gl2010}. From the PQ
and left-right symmetry breaking, we \cite{gl2010} can elegantly
realize the universal \cite{berezhiani1983} seesaw
\cite{minkowski1977} or the double \cite{mohapatra1986} and linear
\cite{barr2003} seesaw.

In this paper we propose a novel left-right symmetric scenario where
the PQ symmetry is realized such that it naturally leads
simultaneously to inflation \cite{ffo1990} and flavor mixing. The
specific model which we will discuss can be embedded in $SO(10)$
GUTs, where the particle content emerges from a bigger picture.
However, we will not discuss this embeddings further and we will
focus on the main mechanism within the left-right framework. At the
left-right level our model contains one Higgs bi-doublet for each
family, two Higgs doublets, two leptoquark doublets and six complex
singlets in the scalar sector while three neutral singlets and three
generations of lepton and quark doublets in the fermion sector. The
six scalar singlets are responsible for a $U(1)^6_{}$ global
symmetry breaking at the Planck scale. Because of the Yukawa
interactions between the scalar and fermion singlets, the
$U(1)^6_{}$ symmetry is explicitly broken down to a $U(1)^3_{}$
symmetry \cite{bhos2005}. Three Nambu-Goldstone bosons (NGBs) will
obtain heavy masses through the Coleman-Weinberg potential
\cite{cw1973} while the other three will pick up tiny masses through
the color anomaly \cite{adler1969}. The heavy and light pNGBs can
act as the inflaton and the axion, respectively. This inflationary
scenario can also avoid the cosmological domain wall problem
\cite{sikivie1982}. In the absence of any off-diagonal Yukawa
couplings involving the lepton and quark doublets, we can make use
of the mixed fermion singlets to induce the lepton mixing by
tree-level seesaw \cite{smirnov1993} and the quark mixing by
one-loop diagrams.

\emph{The Model}: The scalar sector includes one Higgs bi-doublet
for each family,
\begin{eqnarray}
\phi_i^{}(\textbf{1},\textbf{2},\textbf{2}^\ast_{},0)=\left[\begin{array}{cc}\phi_{i1}^{0}&\phi_{i2}^{+}\\
[2mm]\phi_{i1}^{-}&\phi_{i2}^{0}\end{array}\right]\,,
\end{eqnarray}
two Higgs doublets,
\begin{eqnarray}
\hskip -.4cm\begin{array}{cc}
\chi_L^{}(\textbf{1},\textbf{2},\textbf{1},-1)=\left[\begin{array}{l}\chi_L^0\\
[2mm] \chi_L^-\end{array}\right]\,,&
\chi_R^{}(\textbf{1},\textbf{1},\textbf{2},-1)=\left[\begin{array}{l}\chi_R^0\\
[2mm] \chi_R^-\end{array}\right]\,,
\end{array}
\end{eqnarray}
two leptoquark doublets,
\begin{eqnarray}
\hskip -.6cm \begin{array}{ll}
\eta_L^{}(\textbf{3},\textbf{2},\textbf{1},\frac{1}{3})=\left[\begin{array}{l}\eta_L^{+2/3}\\
[-1mm] \eta_L^{-1/3}\end{array}\right],&
\eta_R^{}(\textbf{3},\textbf{1},\textbf{2},\frac{1}{3})=\left[\begin{array}{l}\eta_R^{+2/3}\\
[-1mm] \eta_R^{-1/3}\end{array}\right],
\end{array}
\end{eqnarray}
and six complex singlets,
\begin{eqnarray}
\sigma_{ij}^{}(\textbf{1},\textbf{1},\textbf{1},0)=\sigma_{ji}^{}(\textbf{1},\textbf{1},\textbf{1},0)\,.
\end{eqnarray}
In the fermion sector, there are three neutral fermion singlets,
\begin{eqnarray}
S_{R_i^{}}^{}(\textbf{1},\textbf{1},\textbf{1},0)\,,
\end{eqnarray}
and three generations of quark and lepton doublets,
\begin{eqnarray}
\hskip -.4cm
\begin{array}{l}q_{L_i^{}}^{}(\textbf{3},\textbf{2},\textbf{1},\,~\frac{1}{3})=\left[\begin{array}{c}
u_{L_i^{}}^{}\\
d_{L_i^{}}^{}
\end{array}\right],\\
[5mm]
l_{L_i^{}}^{}(\textbf{1},\textbf{2},\textbf{1},-1)=\left[\begin{array}{c}
\nu_{L_i^{}}^{}\\
e_{L_i^{}}^{}
\end{array}\right],
\end{array}
\begin{array}{l}q_{R_i^{}}^{}(\textbf{3},\textbf{1},\textbf{2},\,~\frac{1}{3})=\left[\begin{array}{c}
u_{R_i^{}}^{}\\
d_{R_i^{}}^{}
\end{array}\right]\,,\\
[5mm]
l_{R_i^{}}^{}(\textbf{1},\textbf{2},\textbf{1},-1)=\left[\begin{array}{c}
\nu_{R_i^{}}^{}\\
e_{R_i^{}}^{}
\end{array}\right]\,.\end{array}
\end{eqnarray}

We assume a discrete left-right symmetry which is connected to charge-conjugation
and the fields thus will transform as
\begin{eqnarray}
\label{lrsymmetry}
\begin{array}{cccc}
\phi_{i}^{}\leftrightarrow\phi_{i}^{T}\,,&\chi_{L}^{}\leftrightarrow\chi_{R}^{\ast}\,,&
\eta_{L}^{}\leftrightarrow\eta_{R}^{\ast}\,,&\sigma_{ij}^{}\leftrightarrow \sigma_{ij}^{}\,,\\
[2mm] q_{L_i^{}}^{}\leftrightarrow q_{R_i^{}}^{c}\,,&
l_{L_i^{}}^{}\leftrightarrow
l_{R_i^{}}^{c}\,,&S_{R_i^{}}^{}\leftrightarrow S_{R_i^{}}^{}\,.
\end{array}
\end{eqnarray}
In the presence of the above left-right symmetry, we can impose a
family symmetry $U(1)_F^{}=U(1)^3_{}$, under which the left- and
right-handed fermion doublets carry an equal but opposite charge for
each family, i.e.
\begin{eqnarray}
\label{fsymmetry1}
(-\delta_{i1}^{},-\delta_{i2}^{},-\delta_{i3}^{})\quad
\textrm{for}\quad l_{L_i^{}}^{}\leftrightarrow
l_{R_i^{}}^c~~\textrm{and}~~ q_{L_i^{}}^{}\leftrightarrow
q_{R_i^{}}^{c}\,.
\end{eqnarray}
We also assign the $U(1)_F^{}$ charges for other fields,
\begin{eqnarray}
\label{fsymmetry2}
\begin{array}{ccl}
(-\delta_{i1}^{},-\delta_{i2}^{},-\delta_{i3}^{})& \textrm{for}&
S_{R_i^{}}^{}\,,\\
[2mm]
(\delta_{i1}^{}+\delta_{j1}^{},\delta_{i2}^{}+\delta_{j2}^{},\delta_{i3}^{}+\delta_{j3}^{})
&\textrm{for}& \sigma_{ij}^{}\,,\\
[2mm] (-2\delta_{i1}^{},-2\delta_{i2}^{},-2\delta_{i3}^{}) &
\textrm{for} &
\phi_{i}^{}\,,\\
[2mm] (0,0,0) &\textrm{for}& \chi_{L,R}^{}\,,~\eta_{L,R}^{}\,.
\end{array}
\end{eqnarray}
We further introduce a global symmetry $U(1)_G^{}$, under which the
fields carry the following quantum numbers,
\begin{eqnarray}
\begin{array}{cccccc}
2&\textrm{for}&\sigma_{ij}^{}\,,\\
[2mm]
1&\textrm{for}&\chi_L^{}\leftrightarrow\chi_R^{\ast}\,,&\eta_L^{}\leftrightarrow\eta_R^{\ast}\,,&
S_{R_i^{}}^{c}\,,\\
[2mm] 0 &\textrm{for}&l_{L_i^{}}^{}\leftrightarrow
l_{R_i^{}}^c\,,&q_{L_i^{}}^{}\leftrightarrow q_{R_i^{}}^{c}\,,&
\phi_i^{}\,.
\end{array}
\end{eqnarray}
Next we specify the allowed Yukawa interactions,
\begin{eqnarray}
\label{lagrangian} \mathcal{L}_Y^{} &=& -y_{q_{i}^{}}^{}
\bar{q}_{L_i^{}}^{}\phi^{}_{i} q_{R_i^{}}^{}
-y_{l_{i}^{}}^{}\bar{l}_{L_i^{}}^{}\phi^{}_i l_{R_i^{}}^{}
 -f_{i}^{}(\bar{l}_{L_i^{}}^{}\chi_L^{}S_{R_i^{}}^{} \nonumber\\
&&+\bar{l}_{R_i^{}}^{c}\chi_R^{\ast}S_{R_i^{}}^{})-h_{i}^{}(\bar{q}_{L_i^{}}^{}\eta_L^{}
S_{R_i^{}}^{}+\bar{q}_{R_i^{}}^{c}\eta_R^{\ast} S_{R_i^{}}^{})
\nonumber\\
&&-\frac{1}{2}g_{ij}^{}\sigma_{ij}^{}\overline{S}_{R_i^{}}^{c}S_{R_j^{}}^{}+\textrm{H.c.}\,.
\end{eqnarray}
For simplicity, we do not write down the full scalar potential where
$\alpha_{ij}^{}\sigma_{ii}^{}\sigma_{jj}^{}\textrm{Tr}\tilde{\phi}_{i}^{\dagger}\phi_{j}^{}
+\beta_{ij}^{}\sigma_{ij}^{2}\textrm{Tr}\tilde{\phi}_{i}^{\dagger}\phi_{j}^{}+\textrm{H.c.}$
should be absent due to the global symmetry $U(1)_G^{}$. Instead, we
only give the part relevant for generating the mixing between the
left- and right-handed leptoquarks,
\begin{eqnarray}
\label{potential2} \mathcal{V}\supset
\lambda_i^{}\sigma_{ii}^{}\eta_L^\dagger
\phi_i^{}\eta_R^{}+\kappa\eta_L^\dagger\chi_L^{}\chi_R^\dagger\eta_R^{}+\textrm{H.c.}\,.
\end{eqnarray}

Clearly, the lepton and quark doublets, the Higgs and leptoquark
doublets, the Higgs bi-doublets, the fermion singlets and the scalar
singlets can, respectively, belong to the $\textbf{16}_{F_i^{}}^{}$,
$\textbf{16}_H^{}$, $\textbf{10}_{H_i^{}}^{}$,
$\textbf{1}_{F_i^{}}^{}$ and $\textbf{1}_{H_{ij}^{}}^{}$
representation in $SO(10)$ GUTs.

\emph{Pseudo Nambu-Goldstone bosons}: Each scalar singlet
$\sigma_{ij}^{}$ has an independent phase transformation to perform
a $U(1)^{6}_{}$ symmetry. However, the six scalar singlets have the
Yukawa interactions with the three fermion singlets [the $g$-terms
in Eq. (\ref{lagrangian})] so that the $U(1)^{6}_{}$ symmetry should
be explicitly broken down to a $U(1)^3_{}$ symmetry. After the six
scalar singlets develop their vacuum expectation values (VEVs),
there will be six NGBs, i.e.
\begin{eqnarray}
\sigma_{ij}^{}=\frac{1}{\sqrt{2}}(f_{ij}^{}+\xi_{ij}^{})e^{i\frac{\varphi_{ij}^{}}{f_{ij}}}_{}\,.
\end{eqnarray}
The fermion singlets then obtain their Majorana masses
\begin{eqnarray}
\tilde{M}_{ij}^{}=\frac{1}{\sqrt{2}}g_{ij}^{}f_{ij}^{}e^{i\frac{\varphi_{ij}^{}}{f_{ij}}}_{}
=M_{ij}^{}e^{i\frac{\varphi_{ij}^{}}{f_{ij}}}_{}\,,
\end{eqnarray}
which will result in a Coleman-Weinberg potential
\cite{cw1973,bhos2005},
\begin{eqnarray}
V=
\frac{1}{32\pi^2_{}}\textrm{Tr}\left[(\tilde{M}\tilde{M}^\dagger_{}
\tilde{M}
\tilde{M}^\dagger_{})\ln\left(\frac{\Lambda^2_{}}{\tilde{M}\tilde{M}^\dagger_{}}\right)\right]\,,
\end{eqnarray}
with $\Lambda$ being the ultraviolet cutoff.

Only three NGBs can exist in the Coleman-Weinberg potential while
the other three can be absorbed by the three fermion singlets. For
example, we can take
\begin{eqnarray}
&&S_{R_i^{}}^{}e^{i\frac{\varphi_{ii}^{}}{2f_{ii}}}_{}\Rightarrow
S_{R_i^{}}^{}~~\textrm{and~then}~~\tilde{M}_{ij}^{}=M_{ij}^{}e^{i\frac{\varphi'^{}_{ij}}{f'^{}_{ij}}}_{}\nonumber\\
&&~~~~\textrm{with}~~
\frac{\varphi'^{}_{ij}}{f'^{}_{ij}}=\frac{\varphi_{ij}^{}}{f_{ij}^{}}-\frac{\varphi_{ii}^{}}{2f_{ii}^{}}-\frac{\varphi_{jj}^{}}{2f_{jj}^{}}
\,.
\end{eqnarray}
Clearly, we have $\varphi'^{}_{ij}=0$ for $i=j$. By taking a
reasonable simplification on the logarithm
\begin{eqnarray}
\ln\left(\frac{\Lambda^2_{}}{\tilde{M}\tilde{M}^\dagger_{}}\right)\simeq
\textrm{constant}=\mathcal{O}(1)\,,
\end{eqnarray}
the Coleman-Weinberg potential can be expanded by
\begin{widetext}
\begin{eqnarray}
\hskip -1.cm V&=
&\frac{1}{32\pi^2_{}}\left\{(M_{11}^2+M_{12}^2+M_{13}^2)^2_{}+(M_{12}^2+M_{22}^2+M_{23}^2)^2_{}+(M_{13}^2+M_{23}^2+M_{33}^2)^2_{}\right.
\nonumber\\
\hskip -1.cm && +2(M_{11}^2M_{12}^2 + M_{12}^2 M_{22}^2 + M_{13}^2
M_{23}^2)+2(M_{11}^2 M_{13}^2 + M_{12}^2 M_{23}^2 + M_{13}^2
M_{33}^2)+ +2(M_{12}^2 M_{13}^2 + M_{22}^2 M_{23}^2 + M_{23}^2
M_{33}^2)\nonumber\\
\hskip -1.cm &&+4M_{11}^{} M_{22}^{} M_{12}^2
\cos\left(\frac{2\varphi'^{}_{12}}{f'^{}_{12}}\right)+ 4M_{11}^{}
M_{33}^{} M_{13}^2
\cos\left(\frac{2\varphi'^{}_{13}}{f'^{}_{13}}\right)+4M_{22}^{}
M_{33}^2 M_{23}^2
\cos\left(\frac{2\varphi'^{}_{23}}{f'^{}_{23}}\right)+8M_{12}^{}M_{13}^{}M_{23}^{}\nonumber\\
\hskip -1.cm
&&\left.\times\left[M_{11}^{}\cos\left(\frac{\varphi'^{}_{12}}{f'^{}_{12}}
+\frac{\varphi'^{}_{13}}{f'^{}_{13}}-\frac{\varphi'^{}_{23}}{f'^{}_{23}}\right)
+M_{22}^{}\cos\left(\frac{\varphi'^{}_{12}}{f'^{}_{12}}
-\frac{\varphi'^{}_{13}}{f'^{}_{13}}+\frac{\varphi'^{}_{23}}{f'^{}_{23}}\right)
+M_{33}^{}\cos\left(\frac{\varphi'^{}_{12}}{f'^{}_{12}}
-\frac{\varphi'^{}_{13}}{f'^{}_{13}}-\frac{\varphi'^{}_{23}}{f'^{}_{23}}\right)\right]\right\}\,.
\end{eqnarray}
\end{widetext}
A combination of $\varphi'^{}_{12}$, $\varphi'^{}_{13}$ and
$\varphi'^{}_{23}$ can have a potential of the form as below,
\begin{eqnarray}
V=\mu^4{}(1\pm \cos\frac{\varphi}{f})~\textrm{with}~
\mu=\mathcal{O}(M_{ij}^{})\,,~f=\mathcal{O}(f_{ij}^{})\,.
\end{eqnarray}
The pNGB $\varphi$ can realize the natural inflation for
$\mu=\mathcal{O}(10^{15}_{}\,\textrm{GeV})$ and
$f=\mathcal{O}(M_{\textrm{Pl}}^{})$ \cite{ffo1990}. The Majorana
masses $M_{ij}^{}$ should be determined by the Yukawa couplings
$g_{ij}^{}=\mathcal{O}(10^{-4}_{})$ for the given symmetry breaking
scales $f_{ij}^{}=\mathcal{O}(M_{\textrm{Pl}}^{})$. Note that the
inflaton in the complex $\tilde{M}_{ij}^{}$ implies spontaneous CP
violation.

On the other hand, the Yukawa interactions (\ref{lagrangian}) mean
that the three NGBs absorbed in the three fermion singlets can have
derivative couplings with the quarks,
\begin{eqnarray}
\mathcal{L}&\supset&-\frac{1}{2f_{ii}^{}}(\partial_\mu^{}\varphi_{ii}^{})
(\bar{q}_{L_i^{}}^{}\gamma^\mu_{}q_{L_i^{}}^{}-\bar{q}_{R_i^{}}^{}\gamma^\mu_{}q_{R_i^{}}^{})\nonumber\\
&=&\frac{1}{2f_{ii}^{}}(\partial_\mu^{}\varphi_{ii}^{})(\bar{u}_{i}^{}\gamma^\mu_{}\gamma_5^{}u_{i}^{}
+\bar{d}_{i}^{}\gamma^\mu_{}\gamma_5^{}d_{i}^{})\,.
\end{eqnarray}
Therefore, the NGBs $\varphi_{ii}^{}$ can obtain their tiny masses
through the color anomaly. Clearly, the pNGBs $\varphi_{ii}^{}$ play
the role of the invisible axion \cite{kim1979,dfs1981} while the
family symmetry $U(1)_F^{}$ is identified with the PQ symmetry.
Benefited from the inflation, our model can escape from the
cosmological domain wall problem \cite{sikivie1982}. Furthermore,
for the PQ symmetry breaking at the Planck scale, we can choose the
initial value of the misalignment angle by the anthropic argument
\cite{weinberg1987,linde1988} to give a desired dark matter relic
density \cite{pi1984,kc2008}.

\begin{figure}
\vspace{2.5cm} \epsfig{file=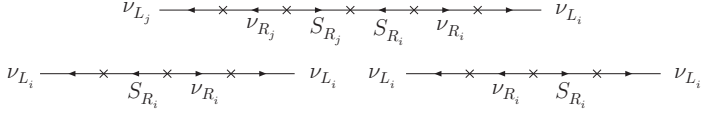, bbllx=6.3cm, bblly=6.0cm,
bburx=16.3cm, bbury=16cm, width=6.cm, height=6.cm, angle=0, clip=0}
\vspace{-7.15cm} \caption{\label{leptonmixing} Tree-level seesaw for
neutrino masses and mixing.}
\end{figure}

\emph{Lepton masses and mixing}: The charged leptons have a diagonal
$3\times 3$ mass matrix, i.e.
\begin{eqnarray}
\label{lepton} \mathcal{L}\supset
-\tilde{m}_e^{}\bar{e}_L^{}e_R^{}+\textrm{H.c.}~~
\textrm{with}~~(\tilde{m}_e^{})_i^{}=y_{l_i^{}}^{}
\langle\phi_{i2}^{0}\rangle\,.
\end{eqnarray}
The mass terms involving neutral leptons are given by
\begin{eqnarray}
\label{lepton2} \mathcal{L}&\supset &
-\tilde{m}_\nu^{}\bar{\nu}_L^{}\nu_R^{}-f\langle\chi_L^0\rangle\bar{\nu}_L^{}S_R^{}
-f\langle\chi_R^{0}\rangle \bar{\nu}_R^c S_R^{}\nonumber\\
&& -\frac{1}{2}\tilde{M}\overline{S}_R^c
S_R^{}+\textrm{H.c.}~\textrm{with}~(\tilde{m}_\nu^{})_i^{}=y_{l_i^{}}^{}\langle\phi_{i1}^{0}\rangle\,.
\end{eqnarray}
For $f\langle\chi_R^{0}\rangle$ an/or $\tilde{M}$ much bigger than
$\tilde{m}_\nu^{}$ and $f\langle\chi_L^0\rangle$, we can make use of
the seesaw formula \cite{minkowski1977} to derive the neutrino
masses, i.e.
\begin{eqnarray}\label{neutrino} \mathcal{L}\supset
-\frac{1}{2}m_\nu^{}\bar{\nu}_L^{}\nu_L^c+\textrm{H.c.}~~
\textrm{with}\quad\quad\quad\quad\quad\quad\quad\quad\quad~&&\nonumber\\
m_\nu^{}=\tilde{m}_\nu^{}\frac{1}{f^T_{}\langle\chi_R^{0}\rangle}\tilde{M}\frac{1}{f\langle\chi_R^{0}\rangle}\tilde{m}_\nu^T
-(\tilde{m}_\nu^{}+\tilde{m}_\nu^T)\frac{\langle\chi_L^0\rangle}{\langle\chi_R^{0}\rangle}\,.&&
\end{eqnarray}
Here the first term is the double seesaw \cite{mohapatra1986} for
$\tilde{M}\gg f\langle\chi_R^{0}\rangle$ or the inverse seesaw
\cite{mv1986} for $\tilde{M}\ll f\langle\chi_R^{0}\rangle$ while the
second term is the linear seesaw \cite{barr2003}, as shown in Fig.
\ref{leptonmixing}. With a left-right symmetry breaking scale
$\langle\chi_R^{0}\rangle=\mathcal{O}(10^{13}_{}\,\textrm{GeV})$,
the double/inverse seesaw term should be the double seesaw for
$\tilde{M}=\mathcal{O}(10^{15}_{}\,\textrm{GeV})$. In this scenario,
the inflaton should decay into the right-handed neutrinos through
the off-shell fermion singlets. Subsequently, the decays of the
right-handed neutrinos can realize the non-thermal or thermal
leptogenesis \cite{fy1986}.

Since the charged lepton mass matrix in Eq. (\ref{lepton}) has a
diagonal structure, the neutrino mass matrix (\ref{neutrino}) should
account for the lepton mixing described by the PMNS matrix. With the
six elements in the Majorana mass matrix $\tilde{M}$, we have enough
flexibility to fit the known masses and mixing \cite{smirnov1993}.

\begin{figure}
\vspace{3.7cm} \epsfig{file=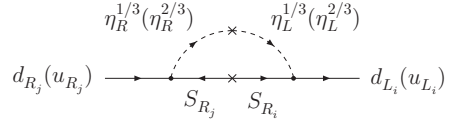, bbllx=5.3cm, bblly=6.0cm,
bburx=15.3cm, bbury=16cm, width=6.cm, height=6.cm, angle=0, clip=0}
\vspace{-8.2cm} \caption{\label{quarkmixing} One-loop diagrams for
quark masses and mixing. }
\end{figure}

\emph{Quark masses and mixing}: At tree level the mass matrices of
the down- and up-type quarks are both diagonal,
\begin{eqnarray}
\label{qmass0} \mathcal{L}&\supset & -\tilde{m}_d^0
\bar{d}_L^{}d_R^{}-\tilde{m}_u^0
\bar{u}_L^{}u_R^{}+\textrm{H.c.}~~\textrm{with}\nonumber\\
&&
(\tilde{m}_d^0)_i^{}=y_{q_i^{}}^{}\langle\phi_{i2}^0\rangle~~\textrm{and}~~(\tilde{m}_u^0)_i^{}=y_{q_i^{}}^{}\langle\phi_{i1}^0\rangle\,.
\end{eqnarray}
At one-loop order the leptoquarks can mediate the mixing of the
fermion singlets to the quark sector, as shown in Fig.
\ref{quarkmixing}.

To calculate the loop-induced quark masses, we define the mass
eigenstates of the leptoquarks,
\begin{subequations}
\begin{eqnarray}
\left\{\begin{array}{l}
\eta_1^{2/3}=\eta_L^{2/3}\cos\theta_{2/3}^{}-\eta_R^{2/3}\sin\theta_{2/3}^{}\,,\\
[2mm]
\eta_2^{2/3}=\eta_L^{2/3}\sin\theta_{2/3}^{}+\eta_R^{2/3}\cos\theta_{2/3}^{}\,,
\end{array}\right.\\
[2mm] \left\{\begin{array}{l}
\eta_1^{1/3}=\eta_L^{1/3}\cos\theta_{1/3}^{}-\eta_R^{1/3}\sin\theta_{1/3}^{}\,,\\
[2mm]
\eta_2^{1/3}=\eta_L^{1/3}\sin\theta_{1/3}^{}+\eta_R^{1/3}\cos\theta_{1/3}^{}\,,
\end{array}\right.
\end{eqnarray}
\end{subequations}
with the rotation angles,
\begin{subequations}
\begin{eqnarray}
\theta_{2/3}^{}\in
[0,\pi]\,\textrm{for}\,\delta_{2/3}^2=0~\textrm{or}~
\theta_{2/3}^{}=\frac{\pi}{4}\,\textrm{for}\,\delta_{2/3}^2\neq 0\,,&&\\
\theta_{1/3}^{}\in [0,\pi]\,\textrm{for}\,\delta_{1/3}^2=
0~\textrm{or}~\theta_{1/3}^{}=\frac{\pi}{4}\,\textrm{for}\,\delta_{1/3}^2\neq
0\,,&&
\end{eqnarray}
\end{subequations}
and the masses,
\begin{subequations}
\begin{eqnarray}
M_{\eta^{2/3}_2}^2-M_{\eta^{2/3}_1}^2=2\delta_{2/3}^2\sin2\theta_{2/3}^{}\,,\\
M_{\eta^{1/3}_2}^2-
M_{\eta^{1/3}_1}^2=2\delta_{1/3}^2\sin2\theta_{1/3}^{}\,.
\end{eqnarray}
\end{subequations}
Here we have defined
\begin{subequations}
\begin{eqnarray}
\delta_{2/3}^2&=&\frac{1}{\sqrt{2}}\lambda_{i}^{}f_{ii}^{}\langle\phi_{i1}^{0}\rangle
+\kappa\langle\chi_L^0\rangle\langle\chi_R^0\rangle\,,\\
\delta_{1/3}^2&=&\frac{1}{\sqrt{2}}\lambda_{i}^{}f_{ii}^{}\langle\phi_{i2}^{0}\rangle\,.
\end{eqnarray}
\end{subequations}
We further rotate the fermion singlets to diagonalize their Majorana
mass matrix,
\begin{eqnarray}
U^\ast_{} \tilde{M} U^\dagger_{}
=\textrm{diag}\{M_{S_1^{}}^{},M_{S_2^{}}^{},M_{S_3^{}}^{}\}
\end{eqnarray}
We then give the loop-induced quark masses as below,
\begin{widetext}
\begin{subequations}
\label{qmass1}
\begin{eqnarray}
(\tilde{m}_d^1)_{ij}^{}&=&\frac{\sin2\theta_{1/3}^{}}{32\pi^2_{}}
h_i^{} h_j^{} U_{ik}^{\dagger}U_{kj}^{T} M_{S_k^{}}^{}
\left(\frac{M_{\eta^{1/3}_2}^2}{M_{\eta^{1/3}_2}^2-M_{S_k^{}}^2}\ln\frac{M_{\eta^{1/3}_2}^2}{M_{S_k^{}}^2}
-\frac{M_{\eta^{1/3}_1}^2}{M_{\eta^{1/3}_1}^2-M_{S_k^{}}^2}\ln\frac{M_{\eta^{1/3}_1}^2}{M_{S_k^{}}^2}\right)
\simeq \frac{\sin^2_{}2\theta_{1/3}^{}}{16\pi^2_{}}
\frac{h_i^{} h_j^{}\tilde{M}_{ij}^{\ast}\delta_{1/3}^2}{M_{\eta^{1/3}_1}^2}\nonumber\\
&=&\left(\frac{\sin2\theta_{1/3}^{}}{1}\right)^2_{}
\left(\frac{h_i^{}h_j^{}}{1}\right)\left(\frac{\tilde{M}_{ij}^{\ast}}{10^{15}_{}\,\textrm{GeV}}\right)
\left(\frac{\delta_{1/3}^2}{10^{19}_{}\textrm{GeV}\cdot
100\,\textrm{GeV}}\right)
\left(\frac{10^{16}_{}\,\textrm{GeV}}{M_{\eta^{1/3}_1}^{}}\right)^2
63.3\,\textrm{GeV}\,,\\
(\tilde{m}_u^1)_{ij}^{}&=&\frac{\sin2\theta_{2/3}^{}}{32\pi^2_{}}
h_i^{} h_j^{} U_{ik}^{\dagger}U_{kj}^{T}M_{S_k^{}}^{}
\left(\frac{M_{\eta^{2/3}_2}^2}{M_{\eta^{2/3}_2}^2-M_{S_k^{}}^2}\ln\frac{M_{\eta^{2/3}_2}^2}{M_{S_k^{}}^2}
-\frac{M_{\eta^{2/3}_1}^2}{M_{\eta^{2/3}_1}^2-M_{S_k^{}}^2}\ln\frac{M_{\eta^{2/3}_1}^2}{M_{S_k^{}}^2}\right)
\simeq \frac{\sin^2_{}2\theta_{2/3}^{}}{16\pi^2_{}} \frac{h_i^{}
h_j^{}\tilde{M}_{ij}^{\ast}\delta_{2/3}^2}{M_{\eta^{2/3}_1}^2}\nonumber\\
&=&\left(\frac{\sin2\theta_{2/3}^{}}{1}\right)^2_{}
\left(\frac{h_i^{}h_j^{}}{1}\right)\left(\frac{\tilde{M}_{ij}^{\ast}}{10^{15}_{}\,\textrm{GeV}}\right)
\left(\frac{\delta_{2/3}^2}{10^{19}_{}\,\textrm{GeV}\cdot
100\,\textrm{GeV}}\right)
\left(\frac{10^{16}_{}\,\textrm{GeV}}{M_{\eta^{2/3}_1}^{}}\right)^2
63.3\,\textrm{GeV}\,.
\end{eqnarray}
\end{subequations}
\end{widetext}
It is easy to check that the tree-level and loop-order quark mass
matrices can induce the desired quark masses and mixing for a proper
parameter choice.

\emph{Summary}: We discussed a left-right symmetric model with PQ
symmetry with the aim to combine in a natural way inflation and
flavor mixing. A key ingredient is the Yukawa interactions between
the six scalar singlets and the three fermion singlets which
explicitly break a $U(1)^6_{}$ symmetry to a $U(1)^3_{}$ symmetry.
Among the six NGBs, three can obtain heavy masses through the
Coleman-Weinberg potential to drive the natural inflation while the
other three can pick up tiny masses through the color anomaly to
solve the strong CP problem and to explain dark matter. Although the
PQ symmetry forbids any off-diagonal Yukawa couplings involving
lepton and quark doublets, we can mediate the mixing of the fermion
singlets to the neutrino sector by tree-level seesaw and to the
quark sector by one-loop diagrams. Our model can be embedded in
$SO(10)$ GUTs, but we did not discuss the details which do not
change the proposed mechanism.

\textbf{Acknowledgement}: PHG is supported by the Alexander von
Humboldt Foundation.

\end{document}